\documentclass[twocolumn,superscriptaddress,longbibliography,aps,preprintnumbers]{revtex4-2}

\usepackage{graphicx,color}
\usepackage{xcolor}
\usepackage{bm,bbm,amsmath,amssymb,braket}
\usepackage{epstopdf}
\usepackage{relsize}
\usepackage{epstopdf}
\usepackage{comment}
\usepackage{soul}
\usepackage{float}
\usepackage{lipsum}
\usepackage[urlcolor=blue,colorlinks=true,citecolor=blue,
	linkcolor=red,pdfstartview={FitH},bookmarks=false]{hyperref}
\usepackage{MnSymbol}
\usepackage{stmaryrd}
\usepackage{mathtools}

\graphicspath{{Figures/}{./Figures/}{.}{./paper_figures/}}
\setstcolor{red}

\definecolor{purple}{rgb}{0.5, 0., 0.8}

\definecolor{persianblue}{rgb}{0.11, 0.22, 0.73}

\begin{document}

\title{Experimental high-dimensional multi-qubit Bell non-locality on a superconducting quantum processor}

\author{Yousef Mafi}\email{Email: yousef.mafi@tuni.fi}
\affiliation{Computational Physics Laboratory, Physics Unit, Faculty of Engineering and Natural Sciences, Tampere University, P.O. Box 692, FI-33014 Tampere, Finland}
\affiliation{Helsinki Institute of Physics P.O. Box 64, FI-00014, Finland}
\author{Ali G.  Moghaddam}\email{Email: ali.moghaddam@aalto.fi}
\affiliation{Computational Physics Laboratory, Physics Unit, Faculty of Engineering and Natural Sciences, Tampere University, P.O. Box 692, FI-33014 Tampere, Finland}
\affiliation{Helsinki Institute of Physics P.O. Box 64, FI-00014, Finland}
\affiliation{Department of Applied Physics, Aalto University, 02150 Espoo, Finland}

\author{Teemu Ojanen}\email{Email: teemu.ojanen@tuni.fi}
\affiliation{Computational Physics Laboratory, Physics Unit, Faculty of Engineering and Natural Sciences, Tampere University, P.O. Box 692, FI-33014 Tampere, Finland}
\affiliation{Helsinki Institute of Physics P.O. Box 64, FI-00014, Finland}

\date{\today}

\begin{abstract}Combining recent advances in superconducting quantum hardware, we explore quantum correlations in a previously inaccessible regime by observing \emph{simultaneously} high-dimensional and many-body Bell non-locality. We report a high-confidence Bell violation in the correlations between two $d=64$-dimensional systems encoded in twelve qubits. For system sizes up to $d=32$, the strength of the observed nonlocal correlations exceeds the quantum upper bound for $d=2$ systems, providing direct evidence of high-dimensional nonlocality. Furthermore, we demonstrate that the observed violation is genuinely collective: all qubits contribute to the nonlocal correlations, while most pairwise correlations across the bipartition remain Bell-local. Our work illustrates how present-day quantum processors enable the exploration of fundamental predictions of quantum mechanics in previously inaccessible regimes and, in turn, how fundamental quantum effects can be used to benchmark their performance.
\end{abstract}

\maketitle

\section{Introduction}

High‑dimensional Bell tests are being actively pursued to 
explore fundamental properties of quantum mechanics and to probe complex forms of entanglement beyond two‑level systems \cite{Mermin1982,CGLMP,Brunner2010PRL,dada2011experimental,Guo2022PRL,Silberhorn2025,miao2026binarisation}. In principle, higher‑dimensional entanglement enables stronger violations of local realism, improved noise tolerance, and access to richer correlation structures that are of fundamental interest in quantum mechanics \cite{Zohren-Gill,Pironio2017PRL}. Beyond foundational interest, high‑dimensional entanglement plays a practical role: it enables denser information encoding, greater robustness against noise, and advantages in quantum information protocols \cite{Zeilinger2000,Gisin2005PRL}. High‑dimensional Bell violations thus serve both as tests of quantum theory and as powerful certification tools for complex experimental systems, especially as quantum technologies scale to larger Hilbert spaces and more sophisticated functionalities \cite{erhard2018twisted,Oxenlowe2019review}.

In this work, we report a robust violation of the Collins--Gisin--Linden--Massar--Popescu (CGLMP) inequality \cite{CGLMP} in the correlations between subsystems of up to six qubits on a programmable superconducting quantum processor. Remarkably, for intermediate subsystem sizes consisting of two to five qubits, our best experimental results for the high-dimensional Bell function exceed the maximum quantum value attainable for $d=2$, providing clear evidence of genuinely high-dimensional nonlocality. 
Moreover, in contrast to photonic implementations, in our approach the high-dimensional system is encoded across multiple qubits, that enables individual qubit-level control within a many-body setting. We exploit this capability to deliberately perturb each qubit and analyze its impact on the observed nonlocality. This demonstrates that the violation is not carried by a small subset of degrees of freedom, but instead arises from the collective structure of the prepared multi-qubit state. 
Finally, we show that although the high-dimensional Bell inequality is violated, most pairwise correlations across the bipartition satisfy the CHSH inequalities. This confirms that the observed nonlocality cannot be attributed solely to pairwise entanglement, but instead requires genuinely multipartite quantum correlations.

The key technical ingredients enabling the observation of a high-dimensional multi-qubit nonlocality are the compact circuit implementation of a Fourier transform by measurement‑conditioned dynamic circuits \cite{Minev2024, Seif2024dynamic} and other operational advances, namely dynamical decoupling (DD) and advanced error mitigation (EM) techniques \cite{Lloyd1999,EM_RMP_2023,M2_Gambetta,Qiskit2024}. Together, these advances demonstrate how incremental technical improvements in quantum computing can open new domains for fundamental exploration. Conversely, the laws of quantum mechanics themselves provide powerful tools for benchmarking quantum devices and certifying the many‑body quantum states they generate. Because the observed nonlocality does not depend on any hardware‑specific details, our results can be regarded as a certification of multipartite entanglement on an error prone but not malicious device. Our work illustrates the reciprocal relationship between technological progress and fundamental discovery: technical advances enable new foundational tests, while foundational principles offer rigorous means of validating quantum hardware.

\section{Results}

\begin{figure*}[t]
    \centering
    \includegraphics[width=1.9\columnwidth]{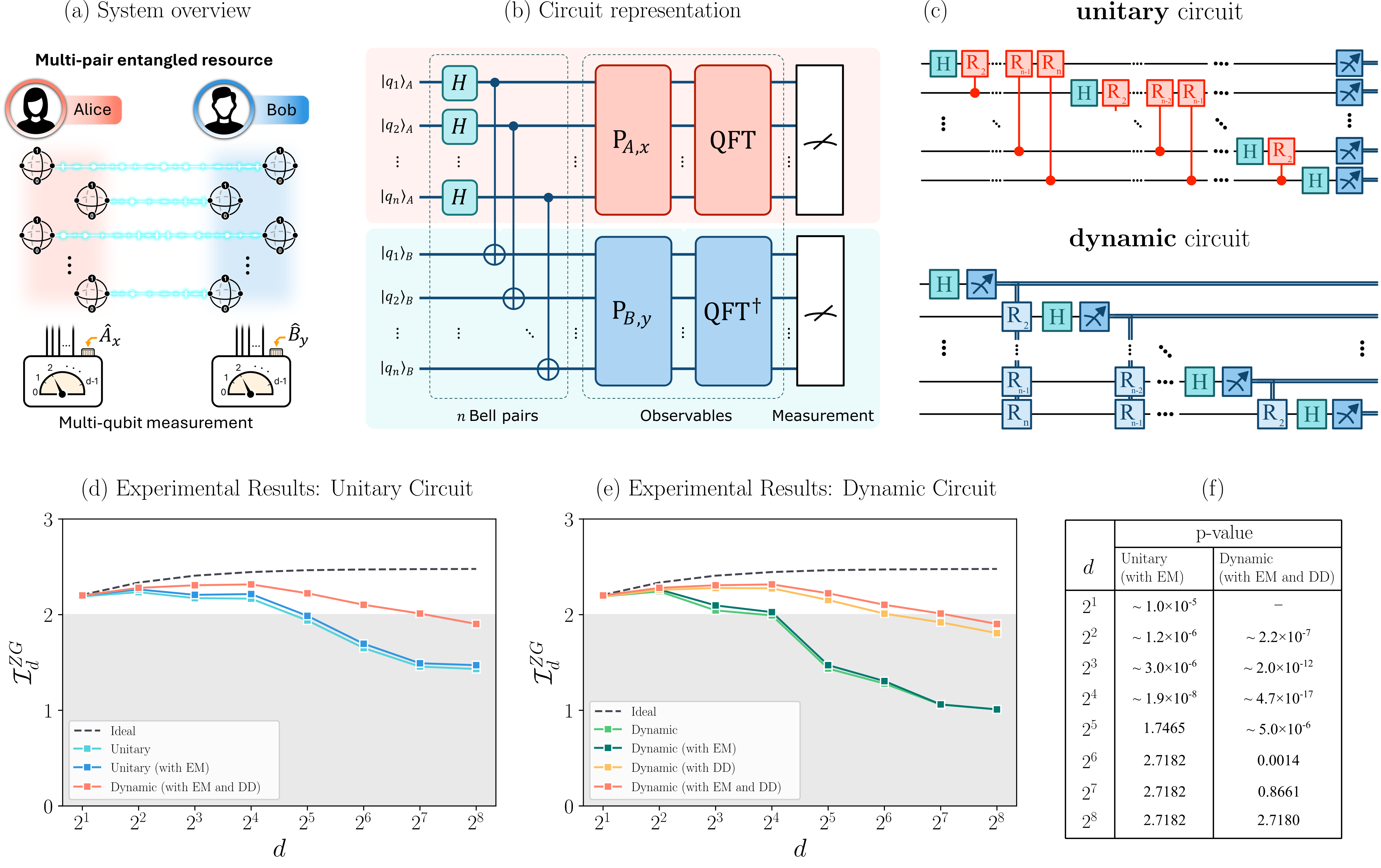}
        \caption{ 
        {\bf High-dimensional non-locality: schematics and results. }
        (a)
        Schematic representation of a high-dimensional Bell test with two parties. (b) Quantum circuit for the multi-qubit Bell test (c) Circuit implementation of a QFT and a DQFT (d) Results for different QFT-based circuits. (e) Results for the DQFT-based circuits. (e) p-values characterizing the likelihood of a LHV description for the experimental results.   }
    \label{fig:1}
\end{figure*}

\subsection{High-dimensional non-locality on superconducting quantum processors}
Typical quantum states in a high-dimensional Hilbert space exhibit a high degree of bipartite entanglement. In contrast, Bell non-locality is far more fragile property, and typical states fail to violate Bell inequalities in practical measurement scenarios. Bell-nonlocal states are those whose correlations cannot be reproduced by any local hidden-variable (LHV) model, meaning that no underlying joint probability distribution over predetermined outcomes can account for the observed statistics. Consequently, the observation of such correlations provides a direct signature of intrinsic nonclassical behavior and offers a means to probe fundamental aspects of quantum mechanics. 

In a bipartite scenario, two parties share a pair of $d$-dimensional systems and independently measure one of two observables, $\hat{A}_1, \hat{A}_2$ and $\hat{B}_1, \hat{B}_2$, as schematically shown in Fig. \ref{fig:1}(a) for a multi-qubit setting. Each measurement yields one of $d$ possible outcomes. Under the assumption of an LHV model, the following CGLMP inequality holds:
\begin{align}\label{eq:Id-ZG}
{\cal I}_d^{\rm ZG} &=  1 - {\mathbb P}_{L}( A_1 < B_1) - {\mathbb P}_{L}( A_2 < B_2)   \nonumber\\
& \qquad  
+{\mathbb P}_{L}(A_2 < B_1 ) 
+ {\mathbb P}_{L}(A_1 \leq B_2)  \leq 2 ,
\end{align}
as first shown by Zohren and Gill \cite{Zohren-Gill}.
The probabilities in Eq.~\eqref{eq:Id-ZG} are defined as
\begin{align}
{\mathbb P}_{L}(A_x < B_y) = \sum_{a=0}^{d-1} \sum_{b=a+1}^{d-1} p(a,b|x,y),
\end{align}
where $p(a,b|x,y)$ denotes the joint probability distribution for obtaining outcomes $a$ and $b$ when measuring observables $\hat{A}_x$ and $\hat{B}_y$ ($x,y = 1,2$) by the two parties, respectively. 
The Bell functional ${\cal I}_d^{\rm ZG}$ is explicitly given in terms of the original $I_d$ introduced in Ref.~\cite{CGLMP}. As shown in Ref. \ref{appndix:A}, the two are related by
\begin{align}\label{eq:Id-IZG}
I_d - 2 = \frac{2d}{d-1} \left({\cal I}^{\rm ZG}_{d} - 2\right).
\end{align}
The violation of the inequality~\eqref{eq:Id-ZG}, or equivalently the condition ${\cal I}^{\rm ZG}_d > 2$, is commonly regarded as a signature of quantumness and the nonclassical nature of the system, as it cannot be explained by LHV theories. Depending on the specific experimental implementation and underlying assumptions, an observed Bell violation may provide partial or even fully device-independent certification of nonclassical behavior. Experimental Bell tests are subject to well-studied loopholes (see, e.g., \cite{Bell2014review,Larsson2014Loopholes,hensen2015loophole,Loophole-Free2015}), which can affect the interpretation and logical implications of a observed violation. In high-dimensional systems, a characteristic challenge is that each measurement must resolve one of the $d$ possible outcomes. Superconducting quantum processors naturally enable genuine $d$-outcome measurements, thereby avoiding the binarization loophole \cite{miao2026binarisation,Binarization2025PRA}, which is a specific concern in high-dimensional Bell experiments, and arises when multilevel outcomes are artificially coarse-grained into binary ones. This issue has posed a particular challenge in optical implementations until very recently \cite{Silberhorn2025,miao2026binarisation}.


States which exhibit robust Bell non-locality in practical measurement scenarios become increasingly rare as the dimension of the system grows, thereby imposing stringent requirements on state preparation in physical platforms. We consider systems composed of $2n$ qubits, which can be viewed as a bipartite system of two $d$-dimensional subsystems with $d = 2^n$. In the computational tensor-product basis, the basis states of each subsystem are written as $|i_1 i_2 \ldots i_n\rangle$, where $i_j \in \{0,1\}$. Each such state can be uniquely labeled by an integer 
$k \in \{0,1,\ldots,d-1\}$ corresponding to its binary representation $k \equiv (i_1 i_2 \ldots i_n)_2$. 

Using this mapping from multi-qubit states to an effective $d=2^n$-dimensional description, and following Ref.~\cite{CGLMP}, we consider the maximally entangled state
\begin{align}\label{eq:maximal-ent}
    |\psi \rangle = \frac{1}{\sqrt{d}} \sum_{k=0}^{d-1} \ket{k}_{A} \otimes \ket{k}_B 
    = \bigotimes_{j=1}^n 
    \Big(
    \frac{1}{\sqrt{2}} \sum_{i_j=0}^1
    \ket{i_j}_{A} \otimes  \ket{i_j}_{B}\Big),
\end{align}
which corresponds to a tensor product of Bell pairs distributed between the two subsystems. The state is measured using two pairs of observables, $\hat{A}_{x} = \hat{U} \hat{\bm P}_{A,x}$ and $\hat{B}_{y} = \hat{U}^\dagger \hat{\bm P}_{B,y}$, applied by the two parties, respectively. A schematic representation of the quantum circuit for state preparation and subsequent measurements is shown in Fig.~\ref{fig:1}(b). The operators $\hat{\bm P}_{A,x}$ and $\hat{\bm P}_{B,y}$ consist of sequences of single-qubit phase gates acting locally on each qubit, while the unitary $\hat{U}$ and $\hat{U}^\dag$ corresponds to a Fourier transform and its inverse (Details of these operators is provided in the Methods section). In practice, this can be implemented via a Quantum Fourier Transform (QFT) on one subsystem and an inverse QFT on the other prior to measurement.

The circuit used for the state preparation and measurement consists of standard operations; however, a naive implementation in larger systems is rapidly degraded by noise. The fidelity of a quantum state typically decreases exponentially in the circuit depth. Deep circuits, such as those generating the states $\ket{\psi}_{x,y} = \hat{A}_x \otimes \hat{B}_y \ket{\psi}$ (i.e., the states prior to computational-basis measurements for different measurement setting $x,y\in \{1,2\}$), are particularly susceptible to noise. This leads to stringent limitations on the system size and on the maximum value of ${\cal I}^{\rm ZG}_d$ that can be observed in practice. The state \eqref{eq:maximal-ent} and the measurement operators $\hat{A}_x$ and $\hat{B}_y$ were originally designed to achieve large values of ${\cal I}^{\rm ZG}_d$. Consequently, even small deviations from ideal gate operations tend to significantly reduce the observed violation. It is therefore essential to optimize circuit performance using all available techniques. 

It was recently shown that a QFT followed by measurements on all qubits can be implemented with a substantial reduction in circuit depth using a dynamic QFT (DQFT) \cite{Minev2024,Seif2024dynamic}. A DQFT circuit, depicted in Fig. \ref{fig:1}(c), employs mid-circuit measurements that condition the subsequent evolution of the circuit. However, this approach introduces additional challenges. In particular, mid-circuit measurements typically require longer execution times than unitary gates; as a result, the total runtime increases, exposing the system to additional noise. To mitigate these effects, we employ dynamical decoupling (DD) \cite{Lloyd1999} and error mitigation (EM) techniques \cite{EM_RMP_2023,M2_Gambetta} available on the IBM platform \cite{Qiskit2024}, as described in Sec.~\ref{QPU} of the Methods.

In Fig. \ref{fig:1}(d)–(e), we present the results obtained from the IBM 156-qubit superconducting quantum processor \texttt{ibm\_aachen}. These figures show the values of the CGLMP parameter ${\cal I}^{\rm ZG}_d$ for different system sizes, obtained using various methods of implementing quantum circuits. In Fig.~\ref{fig:1}(d) and (e), we compare the performance of QFT circuits against the ideal results and the maximum achievable performance of a DQFT implementation. The ideal value of ${\cal I}^{\rm ZG}_d$ grows weakly with the dimensionality of the quantum system and satisfies ${\cal I}^{\rm ZG}_d \leq {\cal I}^{\rm ZG}_{{\rm Q,max}} \approx 2.485$. For system sizes up to four qubits per party, the QFT implementation exhibits a strong violation of the LHV bound, indicating significant Bell non-locality up to $d=16$ level systems; beyond this, ${\cal I}^{\rm ZG}_d$ decreases rapidly. Measurement error mitigation (EM) slightly improves the overall performance but does not yield a qualitative enhancement. The statistical significance of the QFT violation of the LHV bound can be assessed via a $p$-value, which expresses the probability that the observed violation could be consistent with an LHV model. Following Ref. \cite{Wehner2016p-values} (see Sec.~\ref{DATAANALYSIS} in Methods), we calculate upper bounds for the $p$-values. The extremely small values indicate high confidence in the observed CGLMP violation up to $d=16$, effectively ruling out explanations based on LHV models.

The results for DQFT implementations of ${\cal I}^{\rm ZG}_d$ are shown in Fig.~\ref{fig:1}(e). By combining DQFT circuits with dynamical decoupling (DD) and readout EM, we observe significant Bell violations up to six qubits per party, corresponding to Bell non-locality in $d=64$ level systems. Importantly, the experimental values of ${\cal I}^{\rm ZG}_d$ for dimensions $d=2^2$–$2^5$ surpass the corresponding ideal value for ${\cal I}^{\rm ZG}_2$, explicitly signaling high-dimensional non-locality in these cases (corresponding to non-locality between two subsystems each having 2 to 5 qubits). Furthermore, the fact that experimental ${\cal I}^{\rm ZG}_d$ for our best method (dynamic circuits with DD and EM) increases with $d$ up to $d=2^4$ demonstrates that high-dimensional non-locality is stronger and more robust to noise. For larger system sizes, however, noise associated to the increasing circuit depth becomes a limiting factor, and the high-dimensional Bell function decreases on the real hardware for $d \geq 2^5$, even with the best method. 

The associated $p$-values confirm that the above discussed violations are statistically significant. The addition of dynamical decoupling, which suppresses noise arising from the long execution time, is essential for the performance of the DQFT implementation. Measurement EM further improves performance moderately, extending the CGLMP violation from ten shared qubits to twelve. We nominally observe a possible violation even in $d=128$ case (see Sec.~\ref{DATAANALYSIS} for detailed results), however, the violation is weak and the corresponding $p$-value $0.0014$ leaves some room for doubt.  Remarkably, ${\cal I}^{\rm ZG}_d$ is robust to noise: although in the presence of noise, increasing number of qubits and thus circuit depth leads to a progressive (approximately exponential) degradation of process fidelity \cite{Fleckenstein}, the observed difference between ${\cal I}^{\rm ZG}_d$ for the ideal circuit generating $|\psi_{x,y}\rangle$ and the prepared state on real hardware increases only linearly in the studied regime.

\subsection{Revealing collective multipartite non-locality}
In the previous section, we reported violations of the CGLMP inequality for subsystems of up to six qubits ($d = 64$). Notably, for subsystems of up to five qubits, the experimentally obtained values of the CGLMP function, using the most effective circuit implementations, exceed the maximum quantum value achievable for two-qubit systems. To further elucidate the multi-qubit nature of the observed nonlocality, we now investigate how the violation arises from the underlying many-body correlations.

\begin{figure}[t]
\centering\includegraphics[width=0.99\columnwidth]{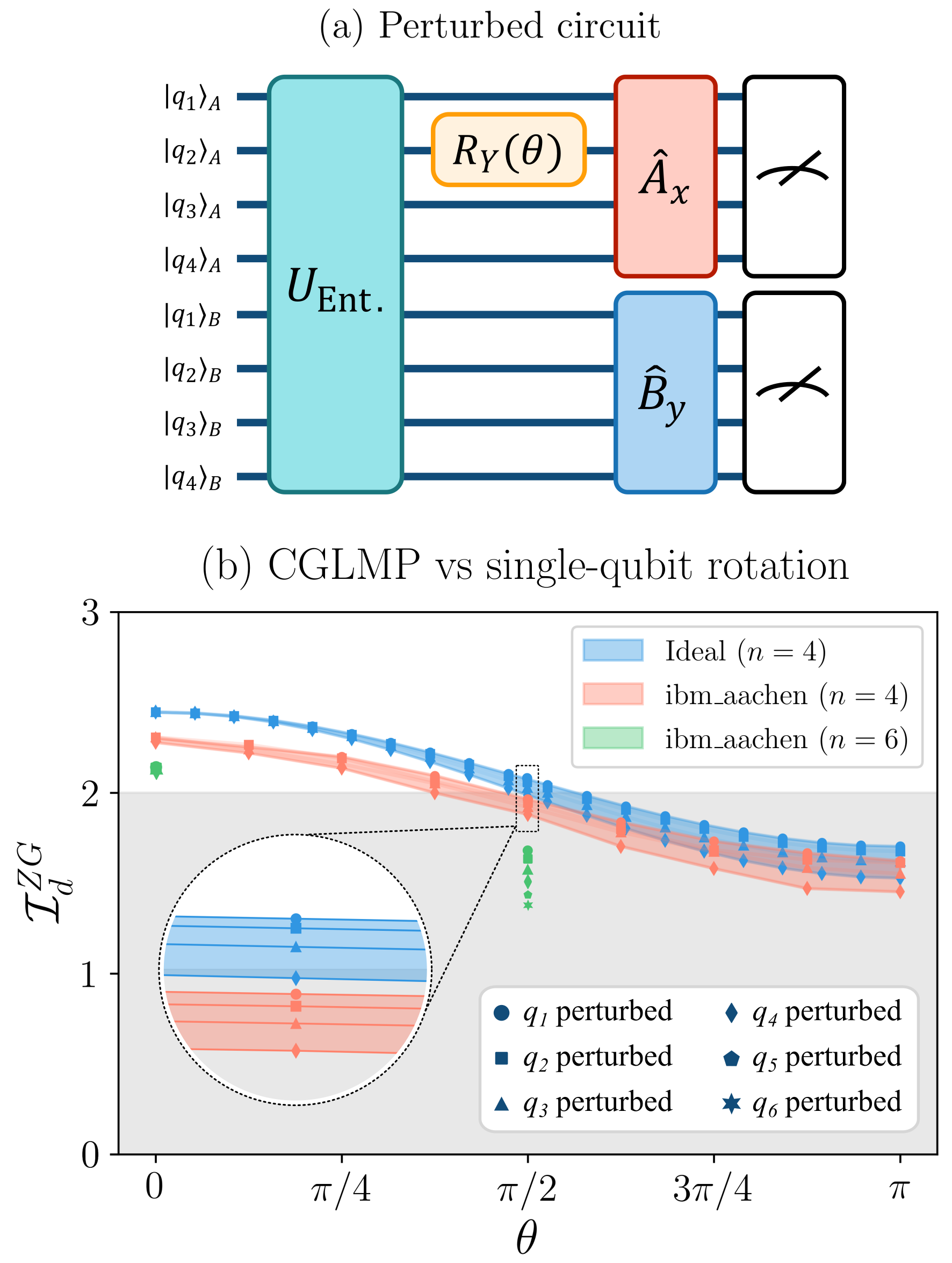}  
        \caption{\textbf{Unraveling the collective nature of nonlocality via single-qubit perturbations} (a) Circuit to probe the participation of an individual qubit to the observed non-locality. Each qubit is individually subjected to a rotation. (b) Evolution of the CGLMP parameter as a function of single-qubit rotations. The blue quadruplet illustrates the effect of rotations of each qubit in a subsystem with four qubits, the red quadruplet illustrates the quantum simulation results.       }
    \label{fig:2}
\end{figure}

We first establish that the observed nonlocality is genuinely collective, in the sense that all qubits in the prepared state actively contribute to the violation. This excludes the possibility that some qubits act merely as passive spectator degrees of freedom. To demonstrate this, it suffices to show that perturbing any individual qubit affects the observed CGLMP violation. This diagnostic is justified by the fact that spectator degrees of freedom could not influence the magnitude of the violation ${\cal I}^{\rm ZG}_d > 2$. 
As a specific single-qubit perturbation, we consider a tilt in the measurement axis of the $i$th qubit by an angle $\theta_i$, as illustrated in Fig.~\ref{fig:2}(a), and evaluate the corresponding CGLMP parameter ${\cal I}^{\rm ZG}_d(\theta_i)$ for different tilt angles. 

The results, shown in Fig.~\ref{fig:2}(b), clearly demonstrate that tilting any individual qubit reduces the observed violation of the LHV bound. 
Moreover, the pattern of sensitivity across different qubits closely follows that of the ideal, noiseless state. In the ideal case, qubits located farther from the bipartition are slightly less sensitive to the tilt; however, the qualitative behavior remains the same across all qubits, indicating that each qubit participates in the observed nonlocality. Even the small variations in how individual qubits affect the value of ${\cal I}^{\rm ZG}_d$ are consistent with the corresponding pattern in the ideal state. These observations therefore highlight the genuinely many-body character of the observed nonlocal correlations.

\begin{figure}
    \centering
    \includegraphics[width=0.99\linewidth]{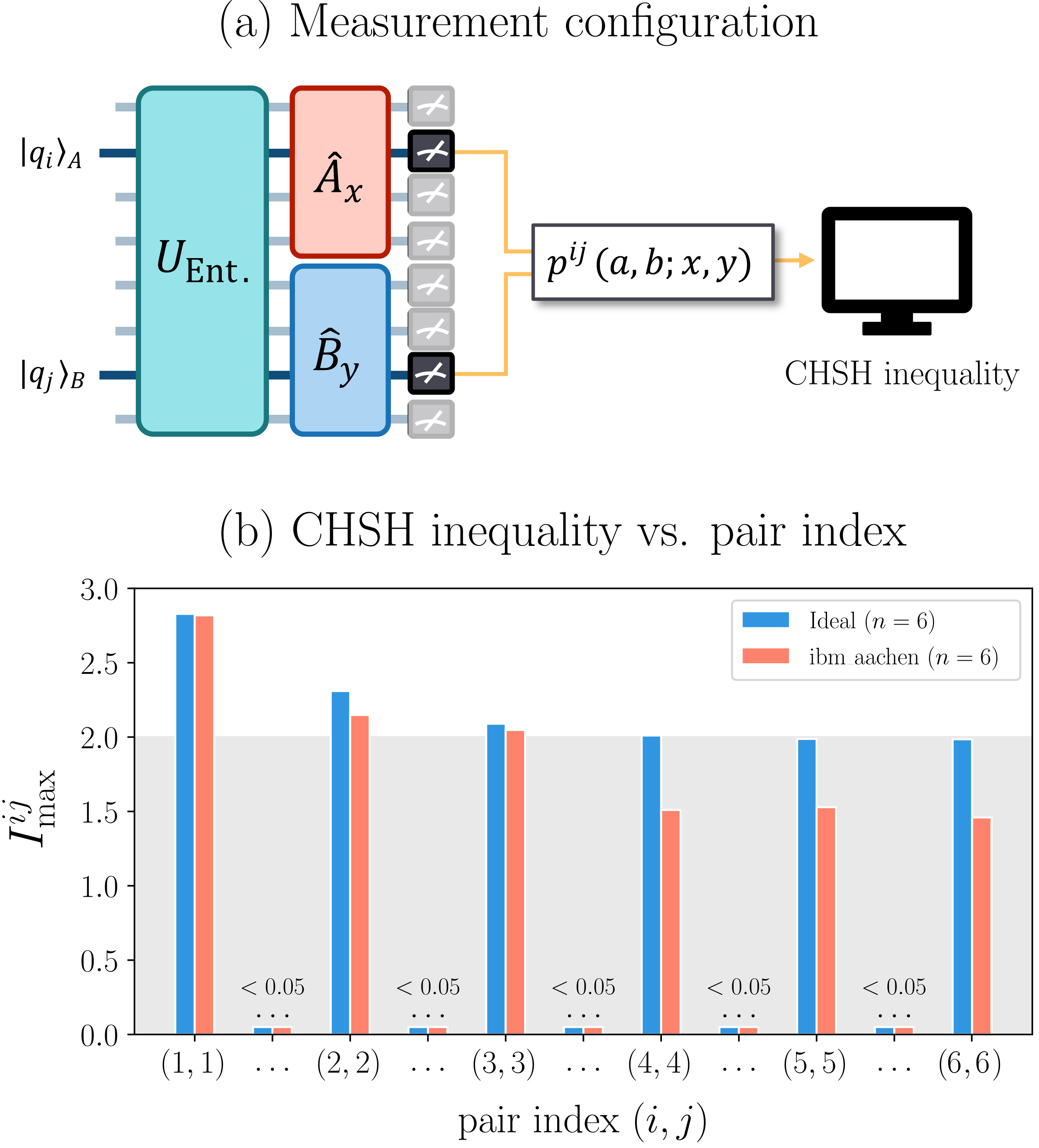}
    \caption{{\bf High-dimensional non-locality against pairwise non-locality} (a) Schematic of the circuit to evaluate pairwise marginal probabilities obtained from the full probability. While the circuit is identical to that in Fig. \ref{fig:1} (a), the measurements results are separately obtained for each pair $i,j$. (b) The pairwise Bell function $I_{\rm max}$ corresponding to CHSH tests. We see only one pair shows siginificant violation while the rest are either slightly violating or Bell local.
    }
    \label{fig:3}
\end{figure}

As a second test, we examine whether the observed nonlocality could arise from a collection of pairwise nonlocal qubit correlations. Ruling out this possibility would imply that the observed violation necessarily originates from genuinely multipartite correlations involving more than two qubits. This can be directly assessed from the measured data. We begin by constructing marginal probability distributions for each qubit pair, $p^{ij}(a,b|x,y)$, from the full probability distribution used to evaluate ${\cal I}^{\rm ZG}_d$ (see Fig. \ref{fig:3}(a) for illustration). Here, $a$ and $b$ denote the outcomes of the $i$th qubit in the first subsystem and the $j$th qubit in the second subsystem, respectively, for measurement settings $x$ and $y$. From these marginals, we compute the correlators
\begin{align}
E_{xy}^{ij} = \sum_{a,b} (-1)^{a+b} p^{ij}(a,b|x,y).
\end{align}
According to Fine's theorem \cite{Fine1982}, a necessary and sufficient condition for the pair $(i,j)$ to admit a LHV model is that the four CHSH inequalities 
\begin{align}
I^{ij}_{\xi\xi'}
&=\left| E_{11}^{ij} + \xi \,E_{12}^{ij} + \xi' \,E_{21}^{ij} - \xi\xi'\,E_{22}^{ij} \right| \le 2,
\end{align}
corresponding to different values of pairs $\xi,\xi' \in \{\pm 1\}$ are satisfied. 
Denoting $I_{\max}^{ij}=\max \{ I_{++}^{ij}, I_{+-}^{ij}, I_{-+}^{ij}, I_{--}^{ij} \}$, this condition can be compactly written as $I_{\max}^{ij} \leq 2$.


As shown in Fig.~\ref{fig:3}(b), the majority of qubit pairs do not violate the LHV bound. In fact, from pair correlations alone, one would be unable to identify the non-locality of half of the qubits in $n=6$ case. This indicates that pairwise nonlocal correlations alone cannot account for the observed CGLMP violation. Instead, genuinely multipartite correlations involving more than two qubits are required to explain the observed nonlocality. Furthermore, the fact that the measured CGLMP parameter ${\cal I}^{\rm ZG}_d$ under individual qubit perturbations follows the qualitative behavior of the ideal, noiseless case suggests that the prepared state faithfully captures the correlation structure of the ideal maximally-entangled state.

\section{Discussion}
The above results demonstrate that current superconducting quantum hardware can support complex many-body states exhibiting both high-dimensional and genuinely multipartite Bell nonlocality. Our findings can be interpreted as the observation of a Bell violation between two $d=64$-dimensional composite systems, in which all degrees of freedom participate in the nonlocal correlations arising from multi-qubit interactions. Recent photonic experiments \cite{miao2026binarisation,Silberhorn2025} that probe genuine $d$-dimensional Bell correlations while closing the binarization loophole have reported violations up to $d=4$ and $d=8$. In comparison, our results extend Bell tests to significantly higher dimensions.

Moreover, high-dimensional nonlocality is not the only remarkable aspect of our work: we also demonstrate that each qubit participates in the observed nonlocal correlations, which cannot be explained by purely pairwise entanglement. While a form of multipartite Bell non-locality, called a Bell depth, has recently been probed for up to 24 qubits on a superconducting processor \cite{Deng2025}, the nature of the Bell depth is fundamentally different from the non-local correlations studied in our work. A large Bell depth arises from relatively low entanglement in ground states of a local Hamiltonian, and both the entanglement structure and the measured observables differ significantly from the ones studied here. Specifically, the approach of Ref.~\cite{Deng2025} reveals multipartite nonlocality by relating the quantum value of particular Bell functionals (evaluated on a given state with suitable observables) to the energy of low-lying states in a Hamiltonian framework \cite{Cirac2017}. Furthermore, due to the locality of the Hamiltonian, multipartite entanglement in such systems is effectively built from an increasing number of pairwise entangled qubits. This type of structure typically does not give rise to high-dimensional nonlocality and violations of the CGLMP inequality. In contrast, our results lie in a previously unexplored regime where high-dimensional Bell nonlocality coexists with a nontrivial multipartite structure: although we observe a clear CGLMP violation at the global level, the non-locality remains mostly hidden from pairwise correlations. This highlights that the observed nonlocality cannot be decomposed into pairwise contributions, but instead emerges from genuinely collective, high-dimensional correlations across the system.

Our results not only shed light on fundamental aspects of quantum mechanics, but also have practical relevance for certifying entanglement in complex many-body systems and benchmarking the performance of quantum hardware. Since the Bell test is performed within a single processor chip, the communication loophole remains open by construction. This inevitably allows for the possibility of signaling correlations, similar to current optical experiments on high-dimensional Bell violations~\cite{miao2026binarisation,Silberhorn2025,Guo2022PRL}. 
Nevertheless, the observed Bell violation provides a stringent and minimally assumption-dependent benchmark of device performance, complementing and in certain respects surpassing standard tomographic methods for entanglement characterization. Unlike tomography, which relies on detailed calibration and noise modeling and becomes exponentially costly in large Hilbert spaces, a many-body Bell test directly probes the global correlation structure generated by the hardware. 

The observed violation demonstrates that the processor produces nonclassical correlations incompatible with any local hidden-variable model, classical noise, or models based solely on pairwise nonlocality. This establishes the high-dimensional Bell test as a robust indicator of genuine multipartite entanglement and coherent collective behavior, and highlights its utility as an alternative to reconstruction-based characterization in scalable multi-qubit systems.
Finally, our work highlights how individual technical innovations in concert can promote the device performance to a higher level, ultimately opening new realms for fundamental exploration. Concretely, we showed how each operational improvement helps to increase the value of the CGLMP parameter, enabling observation of many-body non-locality in substantially larger systems than a straightforward circuit execution would allow.

\section{Summary and outlook}
In this work, we report the observation of genuine high-dimensional, many-body Bell non-locality between two systems consisting of up to six qubits each, thereby extending Bell tests into a new regime. The significance of these results can be understood from two complementary perspectives. On the one hand, we observe multi-qubit non-locality and test fundamental predictions of quantum mechanics in a previously inaccessible regime. On the other hand, our results provide a certification that noisy, yet non-adversarial, quantum hardware can support complex and highly structured many-body states. 

Our work highlights that advances in quantum hardware and control not only enable the exploration of quantum phenomena in new regimes, but also provide general tools for benchmarking device performance. Given that the key technical ingredients enabling this experiment have only recently become available, the prospects for observing high-dimensional many-body non-locality in substantially larger systems in the near future appear promising.

\section{Data availability}
The data that support the findings of this study will be made openly available upon publication.

\section{Code availability}
The quantum simulation code used in this work will be made available upon publication.

\section{Author contributions}
The authors planned the project together. Y.M. carried out the simulations with actual quantum hardware and fake backend, and also performed the data postprocessing. Authors jointly analyzed the results and prepared the manuscript.

\section{Acknowledgments} 
The authors would like to thank Roope Uola for discussions on high-dimensional Bell tests and related topics.
This work was supported by the Finnish Research Council project 362573, the Finnish quantum flagship program and the associated QDOC doctoral pilot program. This work is part of the Finnish Centre of Excellence in Quantum Materials (QMAT). We acknowledge the use of IBM Quantum Credits for this work. The 
views expressed are those of the authors, and do not reflect the official  policy or position of IBM or the IBM Quantum team.

\section{Competing Interests}
The authors declare no competing interests.

\newpage
\section{Methods}

\subsection{CONSTRUCTION OF THE CGLMP MEASUREMENT BASES}
Here, we overview the detailed form of constructing measurement basis described by the unitary operators 
\begin{align}
    \hat{A}_x = \hat{U} \hat{\bm P}_{A,x}, \qquad \hat{B}_y = \hat{U}^\dag \hat{\bm P}_{B,y},
\end{align}
where $x,y \in \{1,2\}$ denoting the labels of observables being measured by Alice and Bob. The first part of these unitaries $\hat{\bm P}_{A,x}$ and $\hat{\bm P}_{B,y}$ involves a set of single-qubit phase operations, whereas the second parts $\hat{U}$ $\hat{U}^\dag$ corresponds to Fourier transforms and its inverse which are implemented through QFT gate operations. 

The phase operators can be expressed as tensor products of single-qubit phase gates,
\begin{align}
    \hat{P}(\varphi)=
    \begin{pmatrix}
    1 & 0 \\
    0 & e^{i\varphi}
    \end{pmatrix}.
\end{align}
Accordingly, the multi-qubit phase operators for Alice and Bob take the form
\begin{equation}
\begin{split}
\hat{\bm P}_{A,x} &= \hat{P}\!\left(\varphi^{A}_{n,x}\right)\otimes \cdots \otimes \hat{P}\!\left(\varphi^{A}_{1,x}\right),\\
\hat{\bm P}_{B,y} &= \hat{P}\!\left(\varphi^{B}_{n,y}\right)\otimes \cdots \otimes \hat{P}\!\left(\varphi^{B}_{1,y}\right),
\end{split}
\end{equation}
where the phase angles applied to the \(j\)-th qubit are given by
$    \varphi^{A}_{j,x} = \pi \, (2^{j}/d) \,\alpha_x$ and
$\varphi^{B}_{j,y} = \pi \, (2^{j}/d) \,\beta_y$,
where $\alpha_1=0$, $\alpha_2=1/2$, $\beta_1=1/4$, and $\beta_2=-1/4$.
Thus the angles depend on the measurement settings $x,y \in \{0,1\}$, and are applied locally to each qubit of Alice and Bob, respectively.

The QFT operator can be formally understood as the discrete Fourier transform over ${\mathbbm Z}_{d=2^n}$:
\begin{align}
    \hat{U} = \frac{1}{\sqrt{d}} \sum_{k =0}^{d-1} \sum_{\tilde{k}=0}^{d-1} e^{2\pi i\: \tilde{k} k/d}  \ket{\tilde{k}} \bra{k}.
\end{align}
Here, the computational basis states $\ket{k}$ correspond to multi-qubit states written as
$\ket{k}= \ket{i_1 i_2 \: \cdots \: i_n}\equiv \ket{i_1} \otimes \cdots \otimes \ket{i_n}$,
where the bit string $i_1 i_2 \: \cdots \: i_n$ represents the binary expansion of the integer $0 \le k \le d-1$.

The implementation of the QFT on an $n$-qubit quantum register can be equivalently expressed as
\begin{align}
    \hat{U}\ket{k} =
    \bigotimes_{j=n}^1   
    \Big[ \ket{0} +  \exp \Big(2\pi i \: 2^{j-1} \frac{k}{d} \Big) \ket{1} \Big].
\end{align}
This transformation can be decomposed into a sequence of unitary operations,
$    \hat{U}= \hat{u}_n  \: \cdots \: \hat{u}_2\: \hat{u}_1$,
where each $\hat{u}_j$ acts on the $j$th qubit. The action of these operators can be written explicitly as
\begin{align}
\hat{u}_{j}\ket{k}
    = \ket{i_1 i_2 \: \cdots \: i_{j-1}} 
    &\otimes 
    \Big[ \ket{0} +  \exp \Big(2\pi i \: 2^{j-1} \frac{k}{d} \Big) \ket{1} \Big] 
    \nonumber\\
 &  
    \otimes
    \ket{i_{j+1} \: \cdots \: i_n}.
\end{align}
Each $\hat{u}_j$ can be implemented using a Hadamard gate $H$ followed by a sequence of controlled phase rotation gates $CR_l$, where $\hat{R}_l = {\rm diag}(1,e^{2\pi i /2^l})$. This construction is illustrated in the upper panel of Fig.~\ref{fig:1}(c).
This decomposition can be understood by substituting $k=\sum_{j'=1}^n i_{j'} 2^{n-j'}$ and observing that each phase factor contributes only when $i_{j'}=1$. This naturally leads to controlled operations conditioned on the states of qubits with indices $j' > j$, i.e., qubits below the $j$th qubit.
Alternatively, the sequence of controlled phase rotations can be implemented as shown in the lower panel of Fig.~\ref{fig:1}(c), using mid-circuit measurements within a dynamic circuit framework as elaborated in details in next part.

\begin{figure}[t!]
    \centering
    \includegraphics[width=0.99\linewidth]{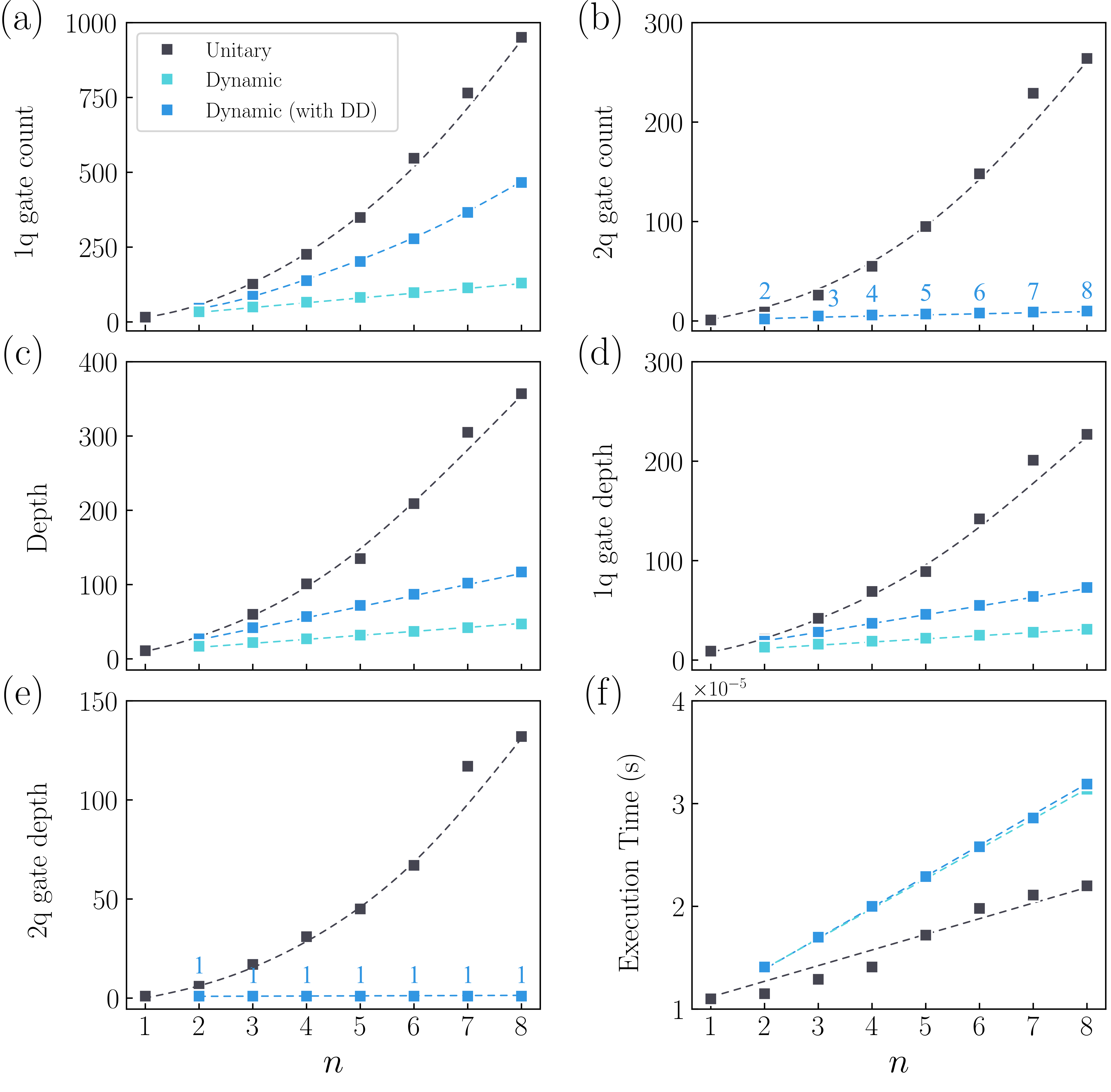}
    \caption{{\bf Resource comparison of multi-qubit Bell test on a quantum processor} (a)-(b) Scaling of single- and two-qubit gate counts, respectively, with system size. (c)-(d) Corresponding circuit depths for each implementation. (e) Execution time, highlighting the overall resource requirements of the protocol.}
    \label{fig:depth}
\end{figure}

\subsection{THE DYNAMIC QUANTUM FOURIER TRANSFORM}\label{DQFT}
We adopted the DQFT framework \cite{Seif2024dynamic} to improve efficiency and scalability. Unlike the standard unitary QFT, which requires $O(n^2)$ two-qubit gates and strict qubit connectivity, DQFT utilizes mid-circuit measurements and classical feed-forward operations to replace coherent controlled gates with classically conditioned operations. This reduces the resource scaling to $O(n)$ while eliminating connectivity constraints.

Figure \ref{fig:depth} illustrates that the unitary approach exhibits quadratic scaling in both single- and two-qubit gate counts as well as circuit depth, leading to rapidly increasing resource demands.  In contrast, the dynamic implementation significantly reduces complexity, achieving near-linear scaling and eliminating most two-qubit gates. However, this comes at the cost of a slight increase in execution time due to the mid-circuit measurement process, as shown in Fig. \ref{fig:depth}(f). The incorporation of DD preserves coherence during idle periods, resulting in improved effective performance without increasing circuit depth. This makes the dynamic approach with DD the most favorable in terms of scalability and noise resilience.

\subsection{SIMULATION DETAILS
ON THE QUANTUM PROCESSOR}\label{QPU}
To perform the quantum simulations, we employed the IBM 156-qubit superconducting quantum processor \texttt{ibm\_aachen} via the IBM Quantum platform \cite{quantumprocessor}. The device utilizes fixed-frequency transmon qubits arranged in a heavy-hex lattice architecture. It demonstrates median coherence times ($T_1$, $T_2$) of approximately $200\  \mu s$, with single- and two-qubit gate fidelities exceeding 99.9\% and 99\%, respectively, and a median readout error of $\sim$ 0.720\%. For benchmarking, high-performance simulators from Qiskit \cite{javadi2024quantum} were used, incorporating realistic noise models that account for gate errors, qubit connectivity, measurement imperfections, and decoherence.

We employed the sampler measurement interface, which executes quantum circuits repeatedly to generate bitstring distributions that faithfully capture raw hardware noise without postprocessing. The number of measurement shots was defined as $m=2\times n\times1024$, scaling with system size.

To reduce noise and decoherence effects, we applied matrix-free measurement mitigation (M3) \cite{M3ref} and dynamical decoupling (DD) \cite{viola1998dynamical, viola1999dynamical, duan1999suppressing}. M3 operates in a reduced subspace defined by observed bitstrings, significantly lowering computational complexity compared to the full Hilbert space. DD enhances coherence preservation during idle periods by applying pulse sequences that refocus qubit states. Specifically, during measurement ($\tau = \tau_{\text{readout}} + \tau_{FF}$), we implemented the 'XY4' sequence with pulse intervals $\tau/4$, $\tau/4$, $\tau/4$, $\tau/4$, $\tau/8$, thereby mitigating decoherence during readout and feedforward operations \cite{Seif2024dynamic}.

\begin{table*}
\caption{
\label{table:1}
{\bf Inequality values across system dimensions} Final values of the inequality ${\cal I}^{\rm ZG}_d$ for increasing system dimension $d$. The table compares ideal results with implementations based on QFT and DQFT circuits, and also  with/without EM and DD. Cases exhibiting a violation in quantum simulations are highlighted in blue, while non-violating cases are shown in red.
The best score is achieved for dynamic circuit implementation with EM and DD.}
\begin{tabular}{|l|p{1.5cm}p{1.5cm}p{1.5cm}p{1.5cm}p{1.5cm}p{1.5cm}p{1.5cm}p{1.5cm}|}
\hline
                         \rule{0pt}{15pt} & \multicolumn{8}{c|}{Inequality value ($I^{ZG}_d$)}                                                  \\[3pt] \cline{2-9} 
                         
\rule{0pt}{15pt} Method                   &\ \ $d=2^1$        & $d=2^2$      & $d=2^3$         & $d=2^4$         & $d=2^5$     & $d=2^6$    & $d=2^7$     & $d=2^8$     \\[3pt] \hline
\rule{0pt}{15pt} Ideal                    &\ \  2.2071& 2.3360& 2.4079 & 2.4457& 2.465     &2.4725    & 2.4762   & 2.4781     \\[3pt]

\rule{0pt}{10pt} Unitary                  &\ \  \textcolor{blue}{2.1909}& \textcolor{blue}{2.2355} & \textcolor{blue}{2.1727}& \textcolor{blue}{2.1672}& \textcolor{red}{1.9394}&\textcolor{red}{1.6507}& \textcolor{red}{1.4572}& \textcolor{red}{1.4315} \\[3pt]
\rule{0pt}{10pt} Unitary (with EM)        &\ \  \textcolor{blue}{2.2010}& \textcolor{blue}{2.2632}& \textcolor{blue}{2.2075}& \textcolor{blue}{2.2153}& \textcolor{red}{1.9875}&\textcolor{red}{1.6958}& \textcolor{red}{1.4919}& \textcolor{red}{1.4720}  \\[3pt]
\rule{0pt}{10pt} Dynamic                  &\ \  -          & \textcolor{blue}{2.2429}& \textcolor{blue}{2.0442}& \textcolor{red}{1.9924}& \textcolor{red}{1.4365}&\textcolor{red}{1.2794}& \textcolor{red}{1.0586}& \textcolor{red}{1.0106} \\[3pt]
\rule{0pt}{10pt} Dynamic (with EM)        &\ \  -          &  \textcolor{blue}{2.2629}& \textcolor{blue}{2.0965}& \textcolor{blue}{2.0263}& \textcolor{red}{1.4722}& \textcolor{red}{1.3055}& \textcolor{red}{1.0617}& \textcolor{red}{1.0093} \\[3pt]
\rule{0pt}{10pt} Dynamic (with DD)        &\ \  -          & \textcolor{blue}{2.2596}& \textcolor{blue}{2.2787}& \textcolor{blue}{2.2751} & \textcolor{blue}{2.1537}& \textcolor{blue}{2.0108}& \textcolor{red}{1.9202}& \textcolor{red}{1.8069} \\[3pt]
\rule{0pt}{10pt} Dynamic (with EM and DD) &\ \  -& \textcolor{blue}{2.2803}& \textcolor{blue}{2.3075}& \textcolor{blue}{2.3167}& \textcolor{blue}{2.2237}& \textcolor{blue}{2.1034}& \textcolor{blue}{2.0117} & \textcolor{red}{1.9030} \\ [3pt] \hline
\end{tabular}
\end{table*}

\subsection{POSTPROCESSED DATA}\label{DATAANALYSIS}

In this section, we provide details on the postprocessing of the output data obtained from the real quantum hardware. The frequency of each measured bitstring is used to estimate the quasi-probabilities of the output state by normalizing the counts with respect to the total number of measurement shots. These quasi-probabilities approximate the underlying outcome distribution; however, due to noise and readout errors, they may not correspond to a valid probability distribution.

To address this issue, we apply the EM-based M3 method, which projects the quasi-probabilities onto the nearest valid probability distribution. This procedure ensures physically consistent probabilities when required. The results are presented in Table \ref{table:1}, which reports the full set of values of the inequality ${\cal I}_d^{\rm ZG}$ for all considered system dimensions. The table compares ideal predictions with both unitary and dynamic implementations and illustrates the impact of EM and dynamical decoupling (DD) across all cases.

As mentioned in the main text, we also compute the $p$-value, which quantifies the statistical significance of the observed violation. Specifically, the $p$-value gives the probability of obtaining results at least as extreme as those measured, under the null hypothesis (i.e., no Bell violation). A smaller $p$-value indicates stronger evidence against the null hypothesis and thus greater confidence in the observed effect. In this work, the $p$-value is used to assess the significance of violations of the ${\cal I}_d^{\rm ZG}$ inequality across different system sizes and implementations. Details of the $p$-value calculation are provided in Appendix \ref{appndix:B}.

\begin{figure}[t!]
    \centering
    \includegraphics[width=0.99\linewidth]{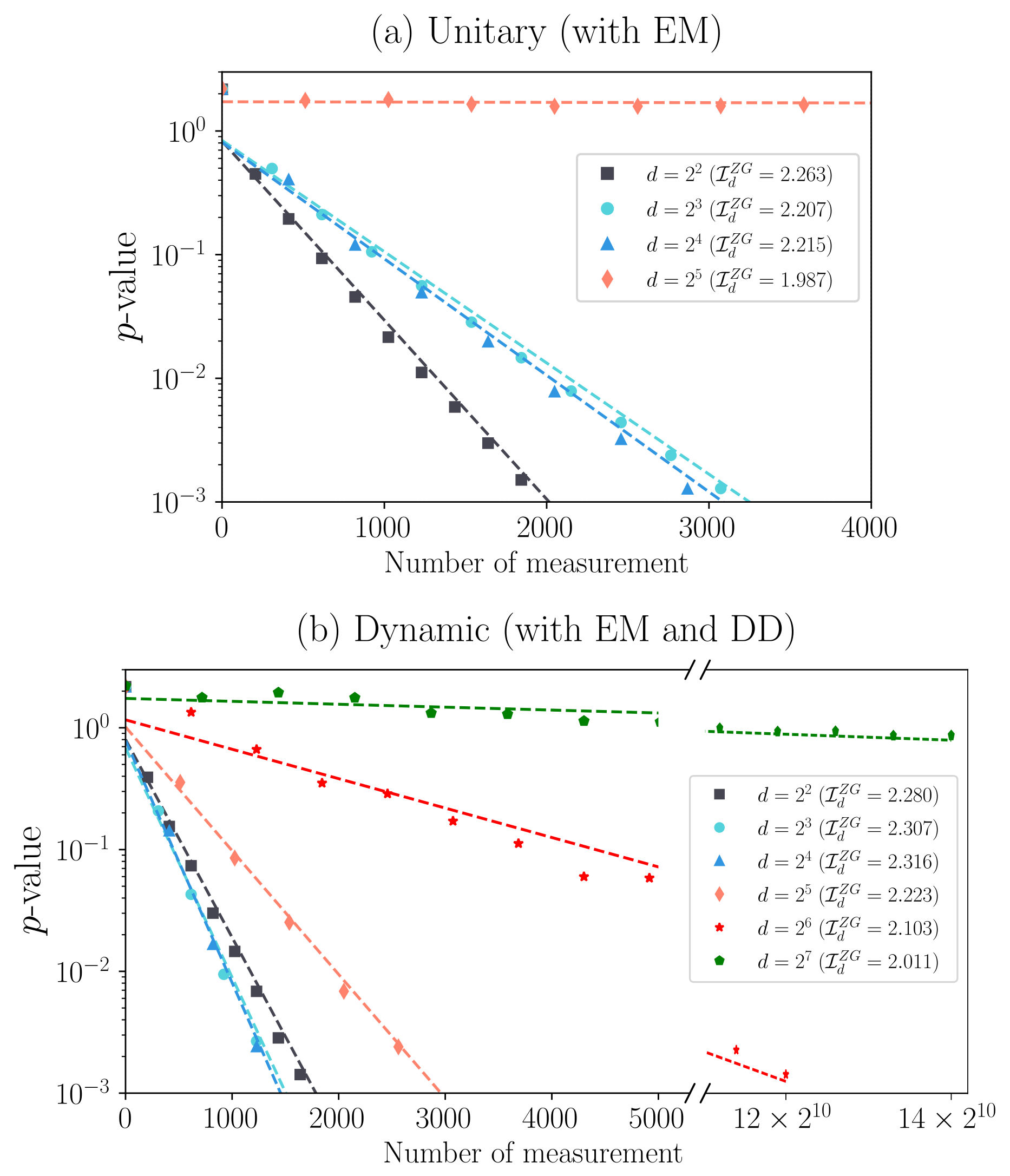}
    \caption{{\bf Statistical significance of ${\cal I}_d^{\rm ZG}$ inequality} $p$-value as a function of the number of attempts (measurements or trials) for different system sizes. (a) Results for QFT-based circuits. (b) Results for DQFT-based circuits.}
    \label{fig:5}
\end{figure}

The dependence of the $p$-value on the number of measurement trials is shown in Fig. \ref{fig:5}(a) and (b) for the unitary and dynamic implementations, respectively. As expected, the $p$-value decreases rapidly with the number of trials, with a faster decay observed for larger violations of ${\cal I}_d^{\rm ZG}$. Conversely, when the Bell parameter approaches the classical bound of $2$, the $p$-value becomes of order unity. This behavior is observed, for example, in the $d = 2^5$ case for the unitary implementation and the $d = 2^7$ case for the dynamic circuit implementation. Finally, the very small $p$-values obtained for subsystems up to dimension $d = 2^6$, together with significant violations ${\cal I}_d^{\rm ZG} - 2 \gtrsim 0.1$, provide strong evidence for high-dimensional nonlocality in the states prepared on real quantum hardware.

\newpage

\newpage

\appendix
\newpage

\section{CGLMP inequalities: derivation and equivalence of different forms}\label{appndix:A}

We begin by recalling the simplified version of the CGLMP inequality, introduced by Zohren and Gill (ZG), which reads
\begin{align}\label{eq:ZG_inequality}
{\mathbb P}_{L}(A_2 < B_2) + {\mathbb P}_{L}(B_2 < A_1) &+ {\mathbb P}_{L}(A_1 < B_1) \nonumber\\
&\quad  + 
{\mathbb P}_{L}(B_1 \leq A_2) \geq 1.
\end{align}
Here, the notation $A_x < B_y$ indicates that the outcome of the measurement $A_x$ is smaller than that of the measurement $B_y$. More explicitly,
\begin{align}
{\mathbb P}_{L}(A_x < B_y) &= \sum_{k=0}^{d-1} \sum_{k'=k+1}^{d-1} p(k,k'|x,y),
\end{align}
and similarly for the relation $A_x \leq B_y$.

The above inequality can be readily proven, as shown in Ref. \cite{Zohren-Gill}, by observing that
\begin{align}
 \{A_2 < B_1\} \subseteq \{A_2 < B_2 \} \cup \{B_2 < A_1\} \cup \{A_1 < B_1\},
\end{align}
which implies
\begin{align}
&{\mathbb P}_{L}(A_2 < B_1) \leq
\nonumber\\
&\qquad{\mathbb P}_{L}(A_2 < B_2) + {\mathbb P}_{L}(B_2 < A_1) + {\mathbb P}_{L}(A_1 < B_1).
\end{align}
Substituting ${\mathbb P}_{L}(A_2 < B_1) = 1 - {\mathbb P}_{L}(B_1 \leq A_2)$ and rearranging the terms yields the ZG form of the CGLMP inequality in Eq.~\eqref{eq:ZG_inequality}.

In order to demonstrating the equivalence between the ZG formulation and the original version of the CGLMP inequality, we first rewrite the ZG expression in a slightly different form more suitable to build the connection to original CGLMP form. Using the identity ${\mathbb P}_{L}(A_x < B_y) = 1 - {\mathbb P}_{L}(B_y \leq A_x)$, the inequality 
\eqref{eq:ZG_inequality}
can be rewritten as
\begin{align}\label{eq:ZG_inequality_2}
{\mathbb P}_{L}(A_2 \leq B_2) 
+1- {\mathbb P}_{L}(A_1 \leq B_2) 
+ {\mathbb P}_{L}(A_1 \leq B_1)  \nonumber\\
+1- {\mathbb P}_{L}(A_2 < B_1) \geq 1 .
\end{align}
By bringing all the probability terms 
to the right hand side, we obtain 
\begin{align}\label{eq:ZG_inequality_modified}
{\cal I}_d^{\rm ZG} &=  1 - {\mathbb P}_{L}( A_1 < B_1) - {\mathbb P}_{L}( A_2 < B_2)   \nonumber\\
& \qquad  
+{\mathbb P}_{L}(A_2 < B_1 ) 
+ {\mathbb P}_{L}(A_1 \leq B_2)  \leq 2 .
\end{align}

\subsection{Equivalence of ZG and CGLMP inequalities}
We now return to the original Bell function introduced in the CGLMP paper, which is given by
\begin{align}\label{eq:Id-GCLMP}
I_d &= 
\sum_{k=0}^{\lfloor d/2 \rfloor -1} \left ( 1 - { 2 k \over d-1} \right) \nonumber\\
& \Bigl( \left[  {\mathbb P}_L(A_1=B_1 + k) +  {\mathbb P}_L(B_1= A_2+k +1) \right.\nonumber\\
& \left . \ \ 
+ {\mathbb P}_L(A_2=B_2 + k ) +
  {\mathbb P}_L(B_2=A_1+ k)\right]\nonumber\\
&  -\left[  {\mathbb P}_L(A_1=B_1 - k-1) +  {\mathbb P}_L(B_1= A_2-k) \right.\nonumber\\
& \left . \  \ 
+ {\mathbb P}_L(A_2=B_2 - k-1 ) +
  {\mathbb P}_L(B_2=A_1- k-1)\right ] \Bigr),
\end{align}
where the equalities are understood modulo $d$, with
\begin{align}
{\mathbb P}_{L}(A_x = B_y + k) &= \sum_{k'=0}^{d-1} p\big(k',k'+k \mod d\:|x,y \big).
\end{align}
Using the periodicity of the modulo operation, individual terms in Eq.~\eqref{eq:Id-GCLMP} can be rewritten. For example, the contribution involving $A_1$ and $B_1$ from the second bracket can be expressed as
\begin{align}\label{eq:first-term}
&\: - \sum_{k=0}^{\lfloor d/2 \rfloor -1} \left ( 1 - { 2 k \over d-1} \right) {\mathbb P}_L(A_1=B_1 + d-k-1) 
\nonumber\\
&=- \sum_{k'=\lfloor d/2 \rfloor}^{d-1} \left ( 1 - { 2 (d-k'-1) \over d-1} \right) {\mathbb P}_L(A_1=B_1 + k') 
\nonumber\\
&=- \sum_{k'=\lfloor d/2 \rfloor}^{d-1} \left ( -1 + { 2 k' \over d-1} \right) {\mathbb P}_L(A_1=B_1 + k') .
\end{align}
This term can then be combined with the corresponding contribution from the first bracket, yielding
\begin{align}
&\sum_{k=0}^{d-1} \left (1 - { 2 k \over d-1} \right) {\mathbb P}_L(A_1=B_1 + k)
\nonumber\\
&\qquad= 1-\sum_{k=0}^{d-1} { 2 k \over d-1} {\mathbb P}_L(A_1=B_1 + k),
\end{align}
where, in the second line, we have used the obvious normalization condition
$
\sum_{k=0}^{d-1} {\mathbb P}_L(A_1=B_1 + k) =1,
$
since this sum accounts for all possible relative outcome differences between $A_1$ and $B_1$. Applying the same procedure to the remaining terms and carrying out straightforward algebraic manipulations, one obtains a more compact form of the original CGLMP expression:
\begin{align}\label{eq:compact-original-CGLMP}
  I_{d}=\sum_{x,y=1}^2\sum_{k=0}^{d-1}\epsilon_{xy}(k)\:{\mathbb P}_{L}(A_{x} = B_{y}+k),
\end{align}
where the coefficients $\epsilon_{xy}(k)$ are given by
\begin{align}
 &\epsilon_{11}(k)=\epsilon_{22}(k)=-\epsilon_{21}(k)=\frac{-2k}{d-1}   \nonumber\\
 &\epsilon_{12}(k)= -2\frac{(k-1) \mod d}{d-1}.
\end{align}

The expression in Eq.~\eqref{eq:compact-original-CGLMP} can be reformulated in terms of expectation values of the differences between outcomes by observing that
\begin{align}
    \sum_{k=0}^{d-1} k\: {\mathbb P}_{L}(A_{x} = B_{y}+k) \equiv {\mathbb E}([A_x-B_y]_d),
\end{align}
where $[X]_d = X - d\lfloor X/d \rfloor$ denotes the value of $X$ modulo $d$. Consequently, the original CGLMP function $I_d$ can be equivalently written as
\begin{align}
    I_d =\frac{-2}{d-1}\: {\mathbb E}\big( 
    &[A_1-B_1]_d + [A_2-B_2]_d
    \nonumber\\
     & - [A_2-B_1]_d- [A_1-B_2-1]_d \big)\nonumber\\
    =\frac{2}{d-1}\: \Big[& d\: {\mathbb E}\big( 
    \lfloor \frac{A_1-B_1}{d} \rfloor 
    +\lfloor \frac{A_2-B_2}{d} \rfloor 
    \nonumber\\
    & - \lfloor \frac{A_2-B_1}{d} \rfloor 
    - \lfloor \frac{A_1-B_2-1}{d} \rfloor \big) -1 \Big] .
\end{align}
Considering that the outcomes of each observable lie in the set $\{0,1,\cdots,d-1\}$, it follows that $\lfloor \frac{A_x-B_y}{d}\rfloor$ is equal to $-1$ if and only if $A_x < B_y$, and $0$ otherwise. Therefore, the corresponding expectation value satisfies
\[
{\mathbb E}\!\left(\lfloor \frac{A_x-B_y}{d}\rfloor\right) = -{\mathbb P}(A_x<B_y).
\]
Substituting this relation into the above expression for $I_d$ yields
\begin{align}
    I_d=\frac{2d}{d-1}\: \Big[& 
     -{\mathbb P}(A_1< B_1)
    -{\mathbb P}(A_2< B_2) \nonumber\\
    & +{\mathbb P}(A_2< B_1)
    +{\mathbb P}(A_1< B_2+1)\Big] -\frac{2}{d-1}\nonumber\\
    =\frac{2d}{d-1}\: \Big[& 
     -{\mathbb P}(A_1< B_1)
    -{\mathbb P}(A_2< B_2) \nonumber\\
    & +{\mathbb P}(A_2< B_1)
    +{\mathbb P}(A_1 \leq B_2)\Big] -\frac{2}{d-1}
    \nonumber\\
    =\frac{2d}{d-1} \big(& {\cal I}^{\rm ZG}_{d}-1\big) -\frac{2}{d-1} \nonumber\\
    = \frac{2d}{d-1} \big(&{\cal I}^{\rm ZG}_{d}-2\big) +2 ,
  \end{align}
which completes the proof of equivalence between 
original and modified forms of CGLMP functions ($I_d$ and ${\cal I}_d^{\rm ZG}$).

\section{Derivation of the $p$-value Computation}\label{appndix:B}

In this section, we present the evaluation of $p$-values, which basically corresponds to the maximum probability that the total score $C$ is at least as large as the observed score $c$, over all LHV models. The $p$-value is computed by formulating the Bell test as a general game, in which each trial contributes a score and the total score serves as the test statistic. To analyze the experimental data, we proceed as follows. In each trial $j$, Alice and Bob choose measurement settings $(x_j, y_j)$ and obtain outcomes $(a_j, b_j)$, which define a score $s_{a_j b_j \mid x_j y_j}$. Summing over all $m$ trials, the total observed score is
\begin{equation}
c = \sum_{j=1}^{m} s_{a_j b_j \mid x_j y_j}.
\end{equation}

To see how the scores are defined, it is useful to view them as a reformulation of a Bell inequality in the generic form
\begin{align}
\beta_{\min} \leq \frac{1}{N}\sum_{x,y}\sum_{a,b} s_{ab | xy} \: p(a,b | x,y) \leq \beta_{\max},
\end{align}
where the coefficients $s_{ab \mid xy}$ define the scoring rule in the game-theoretic picture of a Bell experiment, and $N$ is an appropriate normalization factor. In the case of the CGLMP inequality (Eq.~\ref{eq:Id-GCLMP}), the score function is defined as follows: $s_{ab\mid xy} = +4\left(1-\frac{2k}{d-1}\right)$ if one of the positive conditions is satisfied, $s_{ab\mid xy} = -4\left(1-\frac{2k}{d-1}\right)$ if one of the negative conditions is satisfied, and $s_{ab\mid xy} = 0$ otherwise~\cite{Wehner2016p-values}. Now, we can adapt the scores to the alternative form of the CGLMP inequality, by defining a corresponding score function for ${\cal I}_d^{\rm ZG}$ as
\begin{align}
s_{ab\mid xy} - 2 = \frac{2d}{d-1} \left(s^{\rm ZG}_{ab\mid xy} - 2\right).
\end{align}

Now, as shown in Ref. \cite{Wehner2016p-values}, the $p$-value can be upper bounded using Bentkus’ inequality \cite{bentkus2004hoeffding} as
\begin{equation} 
\begin{split} \text{P-value} \leq e \left( \left(\sum_{i=\lfloor \delta \rfloor}^{m} \binom{m}{i} (\hat{\gamma}_{\max})^i (1-\hat{\gamma}_{\max})^{m-i} \right)^{1-\delta+\lfloor \delta \rfloor} \right. \\ \left. \left(\sum_{i=\lfloor \delta \rfloor}^{m} \binom{m}{i} (\hat{\gamma}_{\max})^i (1-\hat{\gamma}_{\max})^{m-i} \right)^{\delta-\lfloor \delta \rfloor} \right), 
\end{split} 
\end{equation}
where 
\begin{equation} \delta=\sum_{i=1}^{m}\frac{c_i-{s}^{\rm ZG}_{\min}}{{s}^{\rm ZG}_{\max}-{s}^{\rm ZG}_{\min}}, \qquad \hat{\gamma}=\frac{\beta_{\max}-{s}^{\rm ZG}_{\min}}{{s}^{\rm ZG}_{\max}-{s}^{\rm ZG}_{\min}}. 
\end{equation}
Here, $s^{\rm ZG}_{\min}$ and $s^{\rm ZG}_{\max}$ denote the minimum and maximum possible scores, obtained by optimizing over all outcomes and measurement settings $a,b,x,y$. For the CGLMP inequality, $\beta_{\max} = 2$. Furthermore, $c_i$ denotes the score obtained in the $i$th trial. The above bound is used to compute the $p$-values reported in the main text.


\bibliography{bibliography}

\end{document}